\documentclass[conference]{IEEEtran}
\IEEEoverridecommandlockouts
\usepackage{subfigure}
\usepackage{cite}
\usepackage{amsmath,amssymb,amsfonts}
\usepackage{algorithmic}
\usepackage{graphicx}
\usepackage{textcomp}
\usepackage{xcolor}
\usepackage{balance}
\usepackage{verbatim}
\def\BibTeX{{\rm B\kern-.05em{\sc i\kern-.025em b}\kern-.08em
    T\kern-.1667em\lower.7ex\hbox{E}\kern-.125emX}}
\begin{document}

\title{Demo: Low-power Communications Based on RIS and AI for 6G
}

\author{
	\vspace{0.2cm}
	\IEEEauthorblockN{
		Mingyao Cui, Zidong Wu, Yuhao Chen, Shenheng Xu, Fan Yang, and Linglong Dai
		}
	\IEEEauthorblockA{
		Beijing National Research Center for Information Science and Technology (BNRist)\\
		Department of Electronic Engineering,
		Tsinghua University,
		Beijing 100084, China\\
		Emails: \{cmy20, wuzd19, chen-yh21\}@mails.tsinghua.edu.cn,
		\{shxu, fan\_yang, daill\}@tsinghua.edu.cn
	}
}

\maketitle

\begin{abstract}
Ultra-massive multiple-input-multiple-output (UM-MIMO) is promising to meet the high rate requirements for future 6G. However, due to the large number of antennas and high path loss, the hardware power consumption and computing power consumption of UM-MIMO will be unaffordable. To address this problem, we implement a low-power communication system based on reconfigurable intelligent surface (RIS) and artificial intelligence (AI) for 6G. For hardware design, we 
employ a 256-element RIS at the base station to replace the traditional phased array. Moreover, a 2304-element RIS is developed as a relay to assist communication with much reduced transmit power. For software implementation, we develop an AI-based transmission design to reduce computing power consumption. By jointly designing the hardware and software, this prototype can realize real-time 4K video transmission with much reduced power consumption.
\end{abstract}


\section{Themes and Motivation}
Wireless communications have greatly changed people’s lifestyles.
In the future, 6G wireless communications are essential to support new applications, such as meta-verse and holographic video.
Massive multiple-input-multiple-output (MIMO) is the most important technique for current 5G, 
which employs hundreds of antennas to achieve the high data rate of several Gbps.
To achieve a 10-fold increase in data rate for future 6G, massive MIMO is evolving to ultra-massive MIMO (UM-MIMO), where thousands of antennas or even more are employed to improve the spectral efficiency by orders of magnitude \cite{MIMO2_Emil2019}.

One of the key challenges for 6G UM-MIMO, 
is the unaffordable power consumption due to the following two reasons \cite{Overview_Heath2016}.
\begin{itemize}
    \item     Firstly, existing UM-MIMO systems usually employ phased arrays. These phased arrays require a large number of transceiver modules and phase shifters, which results in very high circuit power consumption.
    In addition, due to the high path loss of high-frequency signals, the transmitting power of base station required to achieve reliable communications increases a lot, leading to a high amplifier power consumption. As a consequence, the hardware power consumption of 6G base station is unaffordable. 
    
    \item Secondly, with the substantial increase of the antenna number and required data rate, the computing complexity of traditional signal processing modules also increases significantly, which leads to a high computing power consumption of 6G communication systems in the future.
\end{itemize}

In conclusion, a low-power communication implementation for future 6G is essential.

\section{Scientific and Technical Description}
\begin{figure}
	\centering
	\includegraphics[width=3.3in]{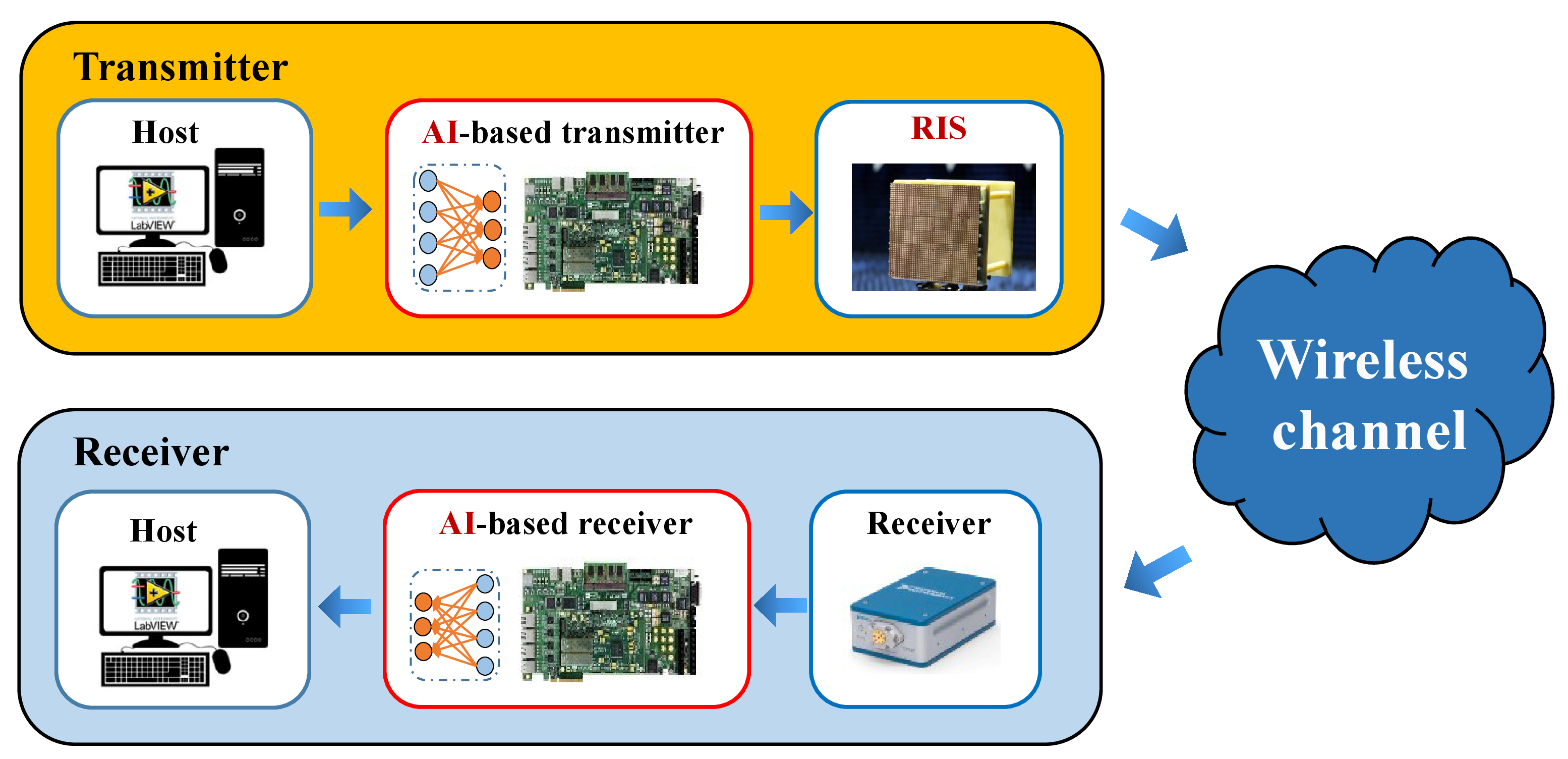}
	\vspace*{-1em}
	\caption{ System setup.
	}
	\label{img:system}
	\vspace*{-1em}
\end{figure}
To address the challenge mentioned above, by jointly designing the hardware and software, we develop a low-power communication system based on reconfigurable intelligent surface (RIS) and artificial intelligence (AI) for 6G as shown in Fig. \ref{img:system}.

\begin{figure}
	\centering
	\subfigure[]
	{\includegraphics[width=2.8in]{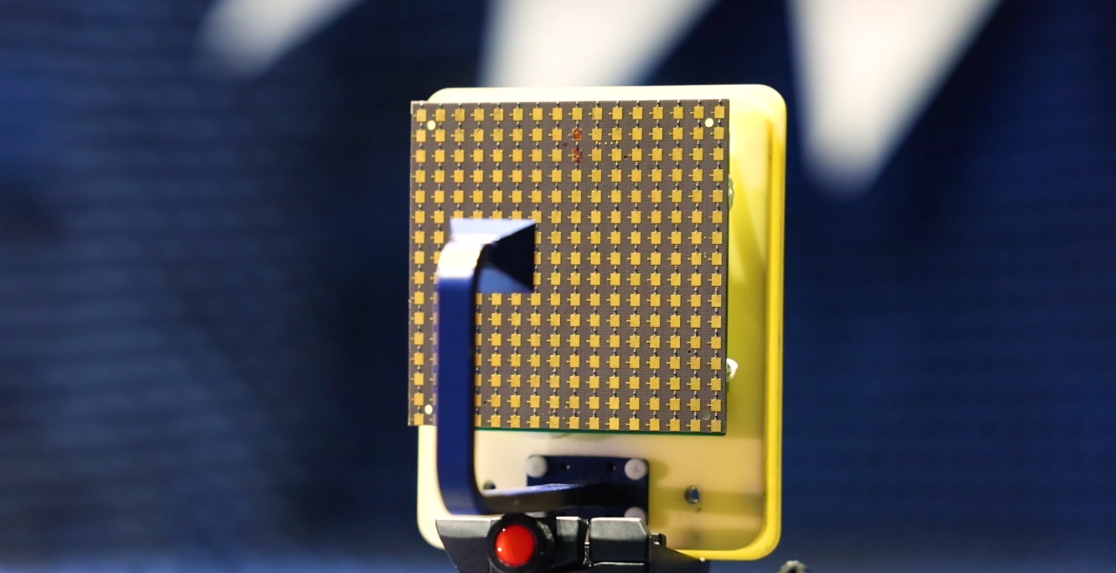}}
	\subfigure[]
	{\includegraphics[width=2.8in]{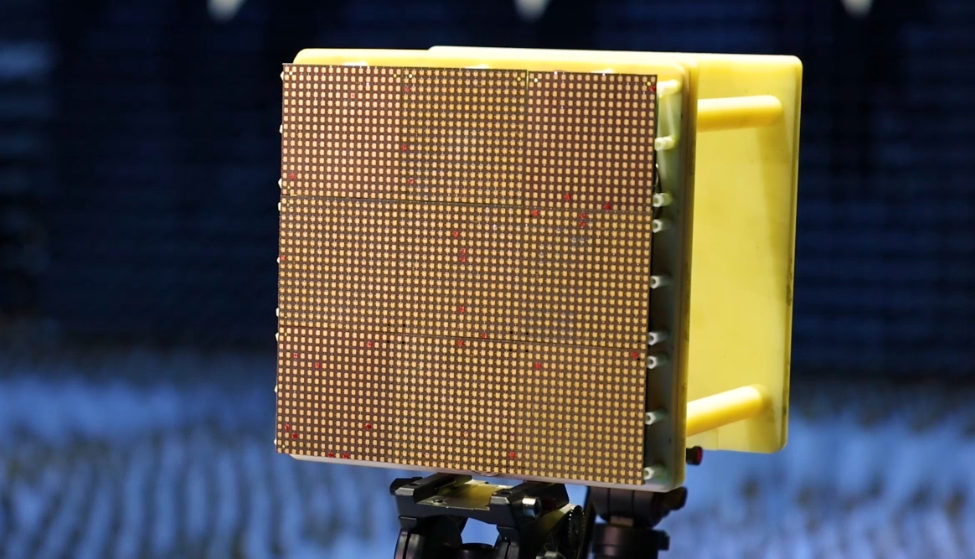}} 
	\\ 
	\centering
	\caption{(a) RIS with 256 elements. (b) RIS with 2304 elements} 
	\vspace*{-1em}
	\label{img:RIS}
\end{figure}
For hardware design, we utilize low-power RIS to replace the phased array. 
RIS integrates the phase shifters and the emission module. It consists of hundreds or thousands of low-power sub-wavelength meta-materials \cite{RISNF_Tang2021, RFocus_Ven21}.
The experiments carried out in our anechoic chamber demonstrate that RIS can generate high-gain beams with low-power consumption by intelligently controlling the electromagnetic waves \cite{activeRIS_Zhang2020}. 
As shown in Fig. \ref{img:RIS} (a), in our system, a RIS with 256 elements is employed at the base station to replace the phased array.
Moreover, as shown in Fig. \ref{img:RIS} (b), a RIS with 2304 elements is employed as a relay to assist the communication with reduced transmit power.

For software design, we develop an AI-based transmission design 
to reduce the computing power consumption. A single neural network-based signal processing module is utilized to replace traditional complex signal processing modules, including modulation, demodulation, channel coding, decoding, and etc. 
Specifically, according to \cite{TCOM}, in the offline training stage, a fully connected neural network is first trained by optimizing the bit error rate (BER) in an end-to-end manner. Then, this neural network is employed in the FPGA modules of the transceiver and receiver for practical online transmission.

\section{Implementation and Contributions}

\begin{figure}
	\centering
	\includegraphics[width=3.5in]{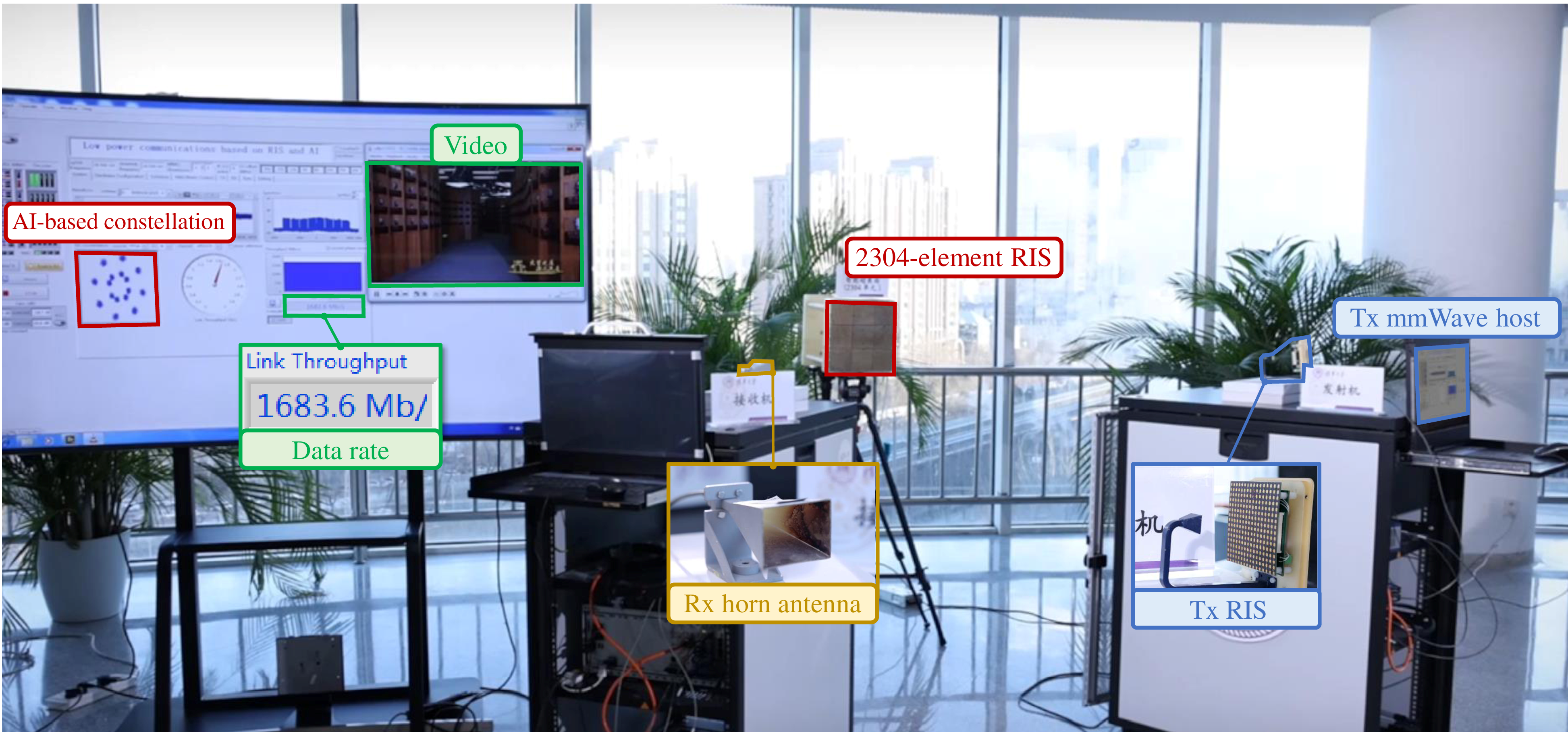}
	\vspace*{-1em}
	\caption{ The developed communication prototype.
	}
	\label{img:system2}
	\vspace*{-1em}
\end{figure}

These two designs have been implemented on our millimeter-wave (mmWave) communication platform to test their performance in real communication environments.
As shown in Fig. \ref{img:system2}, This platform consists of the base station side, the relay side, and the user side. 
The prototype is developed based on NI mmWave Transceiver System (MTS).
The base station and user utilize the software-defined radio (SDR) platform, composed of the PXI hardware architecture and the LabVIEW system, to realize transmission at the mmWave band. 

At the base station, a 4K video stream is first delivered by the transceiver host to the FPGA module. The FPGA module is responsible for the AI-based signal processing, for instance, the transmitted data is mapped to the constellation trained by the neural network. 
Then, the processed data is upconverted to the mmWave band for transmission in wireless channel.
After that, the mmWave signals are radiated to a RIS with 256 elements through a feed source to form a high gain beam. 

At the relay side, a RIS with 2304 elements is utilized to reflect the signals from the base station to the user with a high array gain, so as to reduce the transmit power. 

At the user side, the signals are first received by the Rx horn antenna and then down converted to the base band. Then, real-time data reception is accomplished by a neural network in the receiver FPGA module. 
As shown in Fig. \ref{img:system2}, The receiver host shows the received constellation, the data rate and displays the recovered 4K videos.

The result shows that this prototype can achieve a real-time data rate of 1.68 Gbps. The hardware power consumption can be reduced by 40$\%$, and the computing power consumption is reduced by 20$\%$. 
Based on the prototype, we have also developed cooperation with HUAWEI, ZTE, China Mobile, and British Telecom to promote the application in future 6G systems.

\section{Conclusions}
In this paper, we implemented a low-power communication system based on RIS and AI. This prototype can realize real-time 4K video transmission with much reduced power consumption, which provides a new possibility for future low-power 6G communications.  

\footnotesize
\balance
\bibliographystyle{IEEEtran}
\bibliography{ICC_Demo, IEEEabrv}

\begin{thebibliography}{1}
\providecommand{\url}[1]{#1}
\csname url@samestyle\endcsname
\providecommand{\newblock}{\relax}
\providecommand{\bibinfo}[2]{#2}
\providecommand{\BIBentrySTDinterwordspacing}{\spaceskip=0pt\relax}
\providecommand{\BIBentryALTinterwordstretchfactor}{4}
\providecommand{\BIBentryALTinterwordspacing}{\spaceskip=\fontdimen2\font plus
\BIBentryALTinterwordstretchfactor\fontdimen3\font minus
  \fontdimen4\font\relax}
\providecommand{\BIBforeignlanguage}[2]{{%
\expandafter\ifx\csname l@#1\endcsname\relax
\typeout{** WARNING: IEEEtran.bst: No hyphenation pattern has been}%
\typeout{** loaded for the language `#1'. Using the pattern for}%
\typeout{** the default language instead.}%
\else
\language=\csname l@#1\endcsname
\fi
#2}}
\providecommand{\BIBdecl}{\relax}
\BIBdecl

\bibitem{MIMO2_Emil2019}
E.~Björnson, L.~Sanguinetti, H.~Wymeersch, J.~Hoydis, and T.~L. Marzetta,
  ``Massive {MIMO} is a reality -- what is next? five promising research
  directions for antenna arrays,'' \emph{arXiv preprint arXiv:1902.07678}, Jun.
  2019.

\bibitem{Overview_Heath2016}
R.~W. {Heath}, N.~{González-Prelcic}, S.~{Rangan}, W.~{Roh}, and A.~M.
  {Sayeed}, ``An overview of signal processing techniques for millimeter wave
  {MIMO} systems,'' \emph{IEEE J. Sel. Topics Signal Process.}, vol.~10, no.~3,
  pp. 436--453, Apr. 2016.

\bibitem{RISNF_Tang2021}
W.~{Tang}, M.~Z. {Chen}, X.~{Chen}, J.~Y. {Dai}, Y.~{Han}, M.~{Di Renzo},
  Y.~{Zeng}, S.~{Jin}, Q.~{Cheng}, and T.~J. {Cui}, ``Wireless communications
  with reconfigurable intelligent surface: Path loss modeling and experimental
  measurement,'' \emph{IEEE Trans. Wireless Commun.}, vol.~20, no.~1, pp.
  421--439, Jan. 2021.

\bibitem{RFocus_Ven21}
V.~Arun and H.~Balakrishnan, ``Rfocus: Beamforming using thousands of passive
  antennas,'' in \emph{Proc. 17th USENIX Symposium on Networked Systems Design
  and Implementation (NSDI’20),}, Feb. 2020, pp. 1047--1061.

\bibitem{activeRIS_Zhang2020}
Z.~Zhang, L.~Dai, X.~Chen, C.~Liu, F.~Yang, R.~Schober, and H.~V. Poor,
  ``Active {RIS} vs. passive {RIS}: Which will prevail in {6G}?'' \emph{arXiv
  preprint arXiv:2103.15154}, Mar. 2020.

\bibitem{TCOM}
S.~Cammerer, F.~A. Aoudia, S.~Dörner, M.~Stark, J.~Hoydis, and S.~ten Brink,
  ``Trainable communication systems: Concepts and prototype,'' \emph{IEEE
  Trans. Commun.}, vol.~68, no.~9, pp. 5489--5503, Sep. 2020.

\end{thebibliography}
\end{document}